\documentclass[showpacs,apl,twocolumn,floatfix]{revtex4}

\usepackage{graphicx,float,amsmath}
\usepackage{ mathrsfs }
\usepackage{natbib}

\begin{document}
\date{\today}

\title{Spin and recombination dynamics of excitons and free electrons in p-type GaAs : effect of carrier density}

\author{F.~Cadiz$^{1,2}$}
\author{D. ~Lagarde$^1$}
\author{P. ~Renucci$^1$}
\author{D.~Paget$^2$}
\author{T.~Amand$^1$}
\author{H.~Carr\`ere$^1$}
\author{A. C. H. ~Rowe$^2$}
\author{S.~Arscott$^3$}

\affiliation{%
$^1$Universit\'e de Toulouse, INSA-CNRS-UPS, 31077 Toulouse Cedex, France}

\affiliation{%
$^2$ Laboratoire de Physique de la Mati\`ere Condens\'ee, Ecole P²olytechnique, CNRS, Universit\'e Paris Saclay, 91128 Palaiseau, France}

\affiliation{%
$^3$Institut d'Electronique, de Micro\'electronique et de Nanotechnologie (IEMN), University of Lille, CNRS, Avenue Poincar\'e, Cit\'e Scientifique, 59652 Villeneuve d'Ascq, France}

\begin{abstract}
Carrier and spin recombination are investigated in p-type GaAs of acceptor concentration $N_A=1.5$ x $10^{17}$ cm$^{-3}$ using time-resolved photoluminescence spectroscopy at 15 K. At low photocarrier concentration,  acceptors are mostly neutral and  photoelectrons can either recombine with holes bound to acceptors (e-$\mbox{A}^0$ line) or form excitons which are mostly trapped on neutral acceptors forming the ($\mbox{A}^0$X) complex. It is found that the spin lifetime is shorter for electrons that recombine through the e-$\mbox{A}^0$ transition due to spin relaxation generated by the exchange scattering of free electrons with either trapped or free holes, whereas spin flip processes are less likely to occur once the electron forms with a free hole an exciton bound to a neutral acceptor. An increase of excitation power induces a cross-over to a regime where the bimolecular band-to-band (b-b) emission becomes more favorable 
due to screening of the electron-hole Coulomb interaction and ionization of excitonic complexes and free excitons. Then, the formation of excitons is no longer possible, the carrier recombination lifetime increases and the spin lifetime is found to decrease dramatically with concentration due to fast spin relaxation with free photoholes. In this high density regime, both the electrons that recombine through the e-$\mbox{A}^0$  transition and through the b-b transition have the same spin relaxation time.

\end{abstract}
\pacs{}
\maketitle

Characterization of charge and spin  dynamics in semiconductors is of importance for the designing of devices, such as photovoltaic, microelectronic and spintronics systems. For p-type GaAs, there have been numerous investigations of the photoluminescence spectrum \cite{Scott:1981a,Dingle:1982a,Chen:1994a,feng1995}, as well as of the recombination \cite{nelson1978, seymour1980,horinaka1995,shinohara2000}, spin dynamics \cite{zerrouati1988} and spin-polarized transport \cite{cadiz_prl2013,cadiz2015b}. Among these investigations, the carrier and spin dynamics of p-type material of intermediate doping (around $10^{17}\;\mbox{cm}^{-3}$) considered here has been relatively little investigated \cite{asaka2015, zhao2013}. However, the electron mobility is anticipated to be spin-dependent in this doping range \cite{cadiz2015b}. Moreover, the effect of carrier concentration on this dynamics is still poorly known \cite{amo2007, zhao2009}.

\begin{figure}[htbp]
\includegraphics[clip,width=0.38 \textwidth] {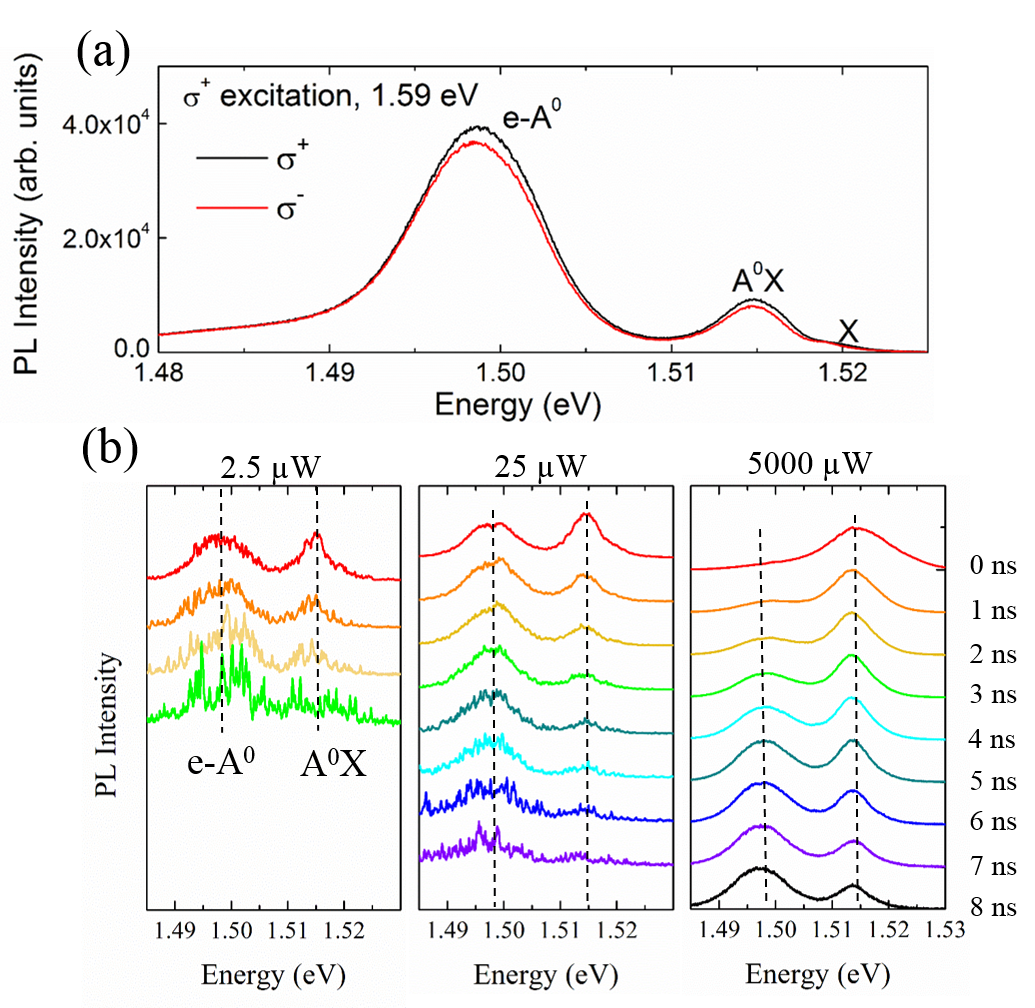}
\caption{a) Steady state polarization-resolved photoluminescence spectrum under circularly polarized excitation at 2 mW, revealing that the two main radiative recombination channels correspond to the e-$\mbox{A}^0$ and to the $\mbox{A}^0$X transitions. b) Transient luminescence spectra at different time delays after the excitation pulse, for a time-averaged excitation power of 2.5 $\mu$W (left panel, where spectra after 3 ns have been omitted because of their weak signal-to-noise), 25 $\mu$W (center panel) and 5 mW (right panel). }
\label{Fig01}
\end{figure}

In the present work, we investigate  charge and spin dynamics as a function of photocarrier density.  These experiments were performed at 15 K  on a 3 $\mu m$ thick GaAs film (Be acceptor concentration $N_A=1.5$ x $10^{17}$ cm$^{-3}$), passivated on both sides by a GaInP layer which confines the photocreated carriers and strongly decreases surface recombination. At low temperatures and for this doping level, the acceptors are mostly neutral \cite{kim1997} and there are two main channels for the electrons to recombine, either with holes bound to acceptor atoms (e-$\mbox{A}^0$ line), or via exciton formation and trapping on acceptors which results in the luminescence of the exciton bound to acceptor line ($\mbox{A}^0$X). We assume that non radiative recombination channels are negligible. 
 These two lines are shown in Fig \ref{Fig01} (a), under a cw $\sigma^+$-circularly polarized excitation at $1.59$ eV, focused onto a beam of  $100 \;\mu m$ diameter and excitation power of $2$ mW. The $\mbox{A}^0X$ line exhibit a small high energy shoulder at $\sim 1.52$ eV corresponding to free exciton recombination (X), identified via temperature-dependent reflectivity and photoluminescence excitation spectroscopy (see supplementary material). The identification of these lines is in agreement with a previous report \cite{Scott:1981a}. Remarkably, the steady-state circular polarization is different for the two transitions, being lower for the e-$\mbox{A}^0$ line ($4 \%$) than for the $\mbox{A}^0$X line ($ 7 \%$).   \\

 We use time-resolved photoluminescence (TRPL) in order to determine the charge and spin lifetimes as a function of excitation power  with a spectral selectivity to $\mbox{A}^0$X or to e-$\mbox{A}^0$ luminescence. As described in Ref. \cite{zhang2013, cadiz2014}, the excitation source was a circularly-polarized mode-locked  Ti:Sa laser ($1.5$ ps pulse width, $80$ MHz repetition frequency, wavelength 780 nm) and the emitted light was dispersed by a spectrometer (resolution 0.12 nm) and detected by a ps streak camera. The time-averaged power was adjusted between 2.5 and 5000 $\mu$W, and since the diameter of the excitation spot (100 $\mu$m) was much larger than the diffusion length, lateral diffusion is negligible. Fig.  \ref{Fig01} b) shows the luminescence spectra at 15 K for different time-delays after the excitation pulse for three different excitation powers, corresponding to values of the photoelectron concentration $n_0$  immediately after the pulse of $6$ x $10^{14}$ cm$^{-3}$  and $1.25$ x $10^{18}$ cm$^{-3}$ for a time averaged excitation power between 2.5 $\mu$W and 5 mW, respectively.  
 A key information obtained from the high energy part of these spectra at high densities is the temperature $T_e$ of the photoelectron gas and its possible dependence on time.  It has been found that while  heating of the electron gas is significant at high power and during the laser pulse, $T_e$ rapidly decreases to a steady value of $50$ K in a time scale of hundreds of ps (see Fig.5 of supplementary material). The hole concentration in the dark $p_0$ is estimated using standard semiconductor statistics \cite{smith1978} to $\sim 10^{16}$ cm$^{-3}$ at 50 K, so that the majority of acceptors are neutral, $N_A^0 \approx 0.9\; N_A = 1.35 \times 10^{17}\;\mbox{cm}^{-3}$. \

As shown in the left panel of Fig. \ref{Fig01}b), for a very weak excitation power, the e-$\mbox{A}^0$ line is dominant at all delays after the pulse. The center panel shows the spectra for a power of 25 $\mu$W, and shows that the two lines have similar intensities immediately after the pulse, while the  e-$\mbox{A}^0$ line again becomes dominant after a delay of about 1 ns. At the highest power excitation of $5$ mW (right panel), exciton formation is completely inhibited  due to screening \cite{Amo:2006a} and the spectrum exhibits mostly band to band (b-b) emission near $1.52$ eV which recombines faster than the e-$\mbox{A}^0$ line due to its bi-molecular nature. This type of crossover has been identified as a Mott transition between an excitonic regime towards an electron-hole plasma regime due to screening of the electron-hole interaction \cite{Collet:1984a,Amo:2006a}, and to our knowledge has never been reported so far in p-type material.  An independent confirmation of exciton screening at high densities can be obtained by analyzing the energy-resolved circular polarization around the $\mbox{A}^0$X /b-b line as a function of excitation power (see Fig.3 of the supplementary material).

\begin{figure}[tbp]
\includegraphics[clip,width=0.43 \textwidth] {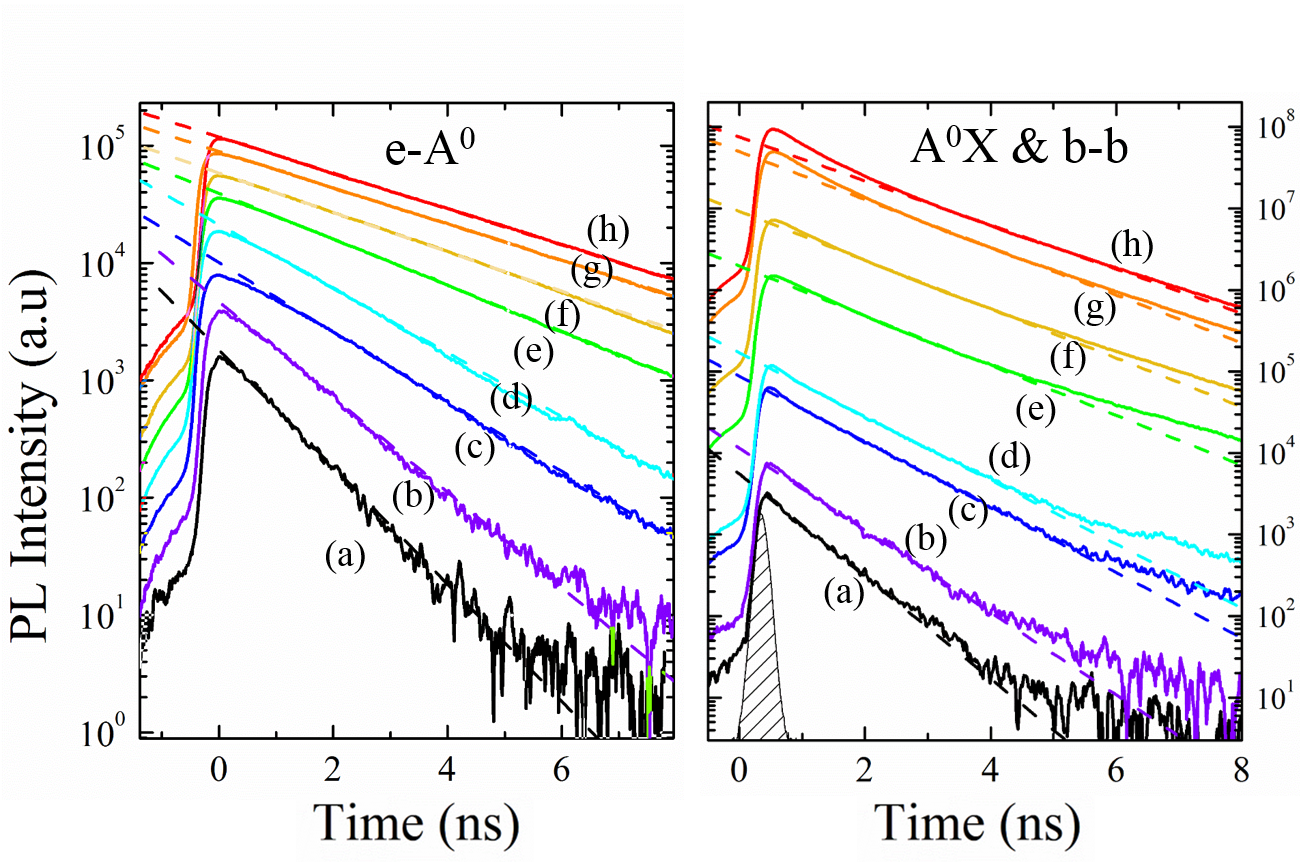}
\caption{Normalized intensity decay transients for the e-$\mbox{A}^0$ line (left panel) and for the $\mbox{A}^0$X line which merges with b-b recombination at high power (right panel). The time-averaged  excitation power for each transient is equal to 2.5 $\mu$W (a), 5 $\mu$W (b), 25 $\mu$W (c), 50 $\mu$W (d), 250$\mu$W (e), 500$\mu$W (f), 2.5 mW (g) and 5 mW (h). The curves have been shifted for clarity. Also shown is the response of the setup (filled curve in the right panel). }
\label{Fig02}
\end{figure}

\begin{figure}[tbp]
\includegraphics[clip,width=0.43 \textwidth] {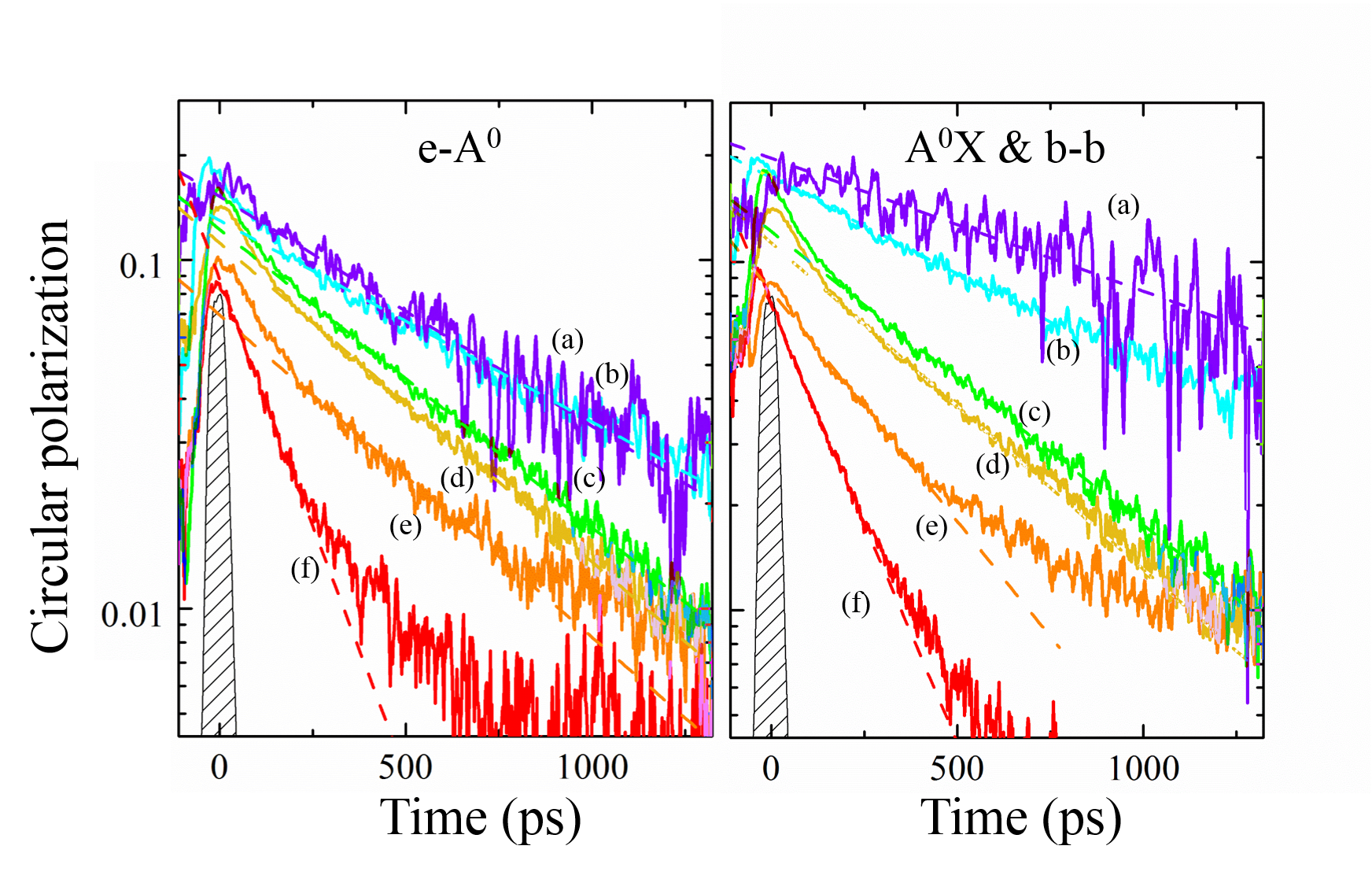}
\caption{ Decay transients of the luminescence degree of circular polarization for the e-$\mbox{A}^0$ line (left panel) and for the $\mbox{A}^0$X line (right panel), for a time-averaged excitation power equal to 5 $\mu$W (a), 50 $\mu$W (b), 250 $\mu$W (c), 500 $\mu$W (d), 2.5 mW (e), and 5 mW (f). These transients directly give the spin relaxation time $T_1$, as measured on the two lines. }
\label{Fig03}
\end{figure}

The intensity transients integrated over each spectral line are summarized in Fig.  \ref{Fig02}; the degree of circular polarization of each line, defined as $\mathscr {P} = (I^{\sigma^+} - I^{\sigma^-})/(I^{\sigma^+}+ I^{\sigma^-})$, where $I^{\sigma^\pm}$ represents the $\sigma^{\pm}$-polarized component of the luminescence, is also shown in Fig. \ref{Fig03}. These transients give directly the photoelectron recombination and spin relaxation time. \

\begin{figure}[htbp]
\includegraphics[clip,width=0.35 \textwidth] {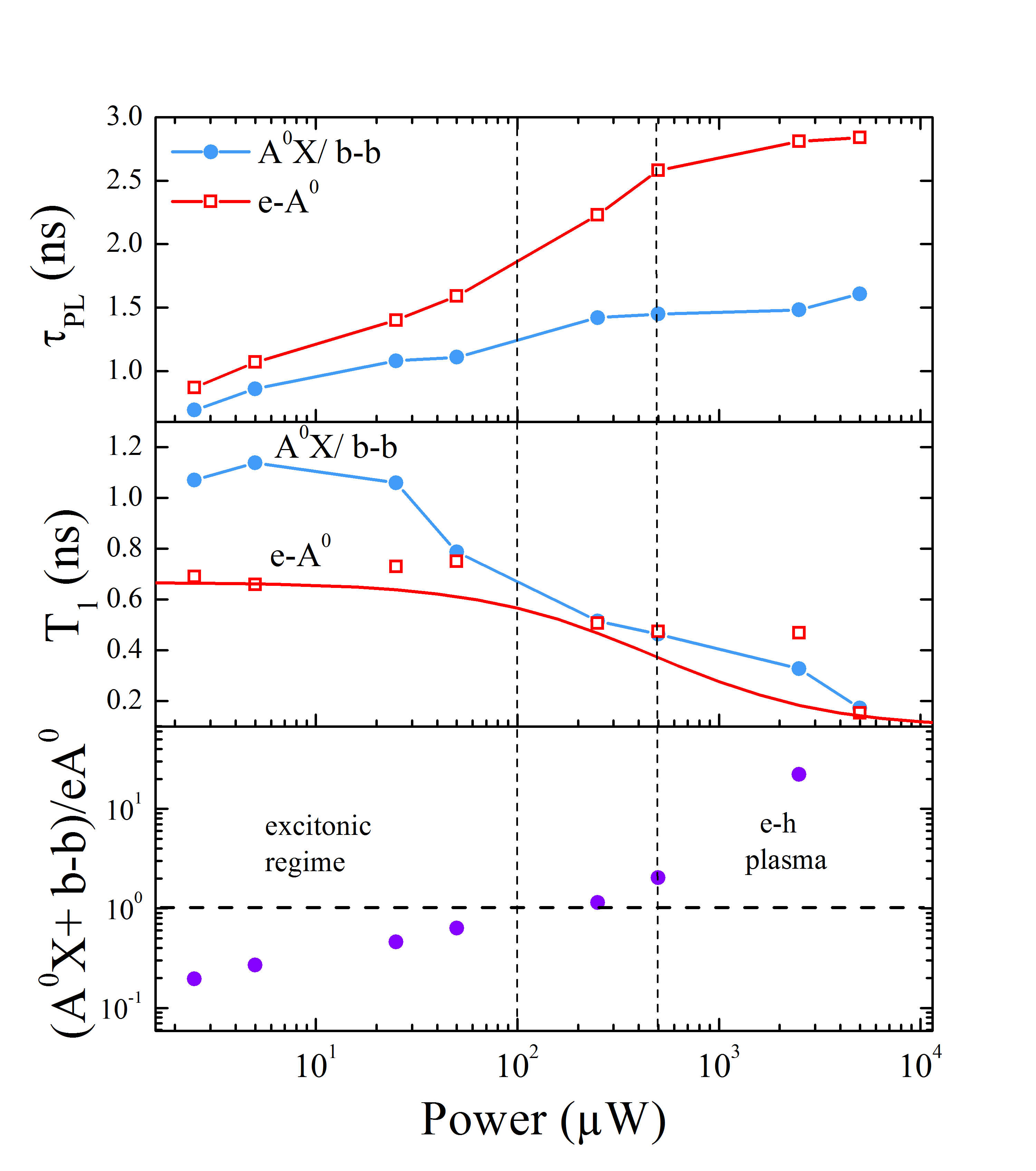}
\caption{The top panel shows the power dependence of the decay times for the two lines, and the middle panel shows the spin relaxation time  found for the two lines, as a function of excitation power. The bottom panel shows the ratio of the integrated intensities for the higher energy line ($\mbox{A}^0$X + b-b) to the e-$\mbox{A}^0$ line at $t=0$, and reveals the existence of a crossover at an excitation power of  about 100 $\mu$W. }
\label{Fig04}
\end{figure}

We first discuss the lifetime $\tau_{PL}$ of both lines as a function of excitation power. 
At low densities, both lines are characterized by a similar decay time of $\tau_{A^0X} = 0.7 \pm 0.05$ ns and $\tau_{e-A^0}=0.9 \pm 0.05 $ ns for $\mbox{A}^0$X and e-$\mbox{A}^0$, respectively.
 In contrast, at the highest excitation density, the e-$\mbox{A}^0$ line decays with $\tau_{e-A^0}=2.8$ ns whereas the b-b recombination has a lifetime of almost half of that of e-$\mbox{A}^0$ line, $\tau_{b-b}=1.6$ ns. The extracted lifetimes  are shown in the top panel of Fig. \ref{Fig04}, where it can be seen that the recombination time for both lines is a monotonic increasing function of the excitation power.
The observed behaviour has a simple interpretation.  At low densities (excitonic regime),  since at short time-delays both e-$\mbox{A}^0$ and $\mbox{A}^0$X lines are present in the spectrum, we assume that during the laser pulse and the subsequent energy relaxation and thermalization, two populations are created, consisting of free electron-hole pairs and of excitons bound to acceptors, respectively. We assume that once the process of thermalization is over, both populations evolve independently as a function of time. For the  e-$\mbox{A}^0$ emission, the luminescence intensity is proportional to $K_0 N_A^0 n$ where $n$ is the photoelectron concentration, $N_A^0$ is the concentration of neutral acceptors and $K_0$ is the bimolecular recombination coefficient between a free electron and a hole bound to acceptor.   We assume for simplicity that $n$ is homogeneous as a function of depth, and this  concentration is a solution of $ \partial n /\partial t= g - K_0 N_A^0 n - n/\tau_{fX} $,
 where $g$ is the number of electrons created in the conduction band per unit volume and per unit time, and $\tau_{fX}$ is the time of exciton formation, which we introduce in order to account for the relatively short lifetime of the e-$\mbox{A}^0$ emission \cite{notetrpl}.  Indeed, using Ref. \cite{dumke1963} we calculate an acceptor recombination time of $1/(K_0 N_A^0)=15 $  ns, that is more than one order of magnitude larger than the measured lifetime of the e-$\mbox{A}^0$ line. We conclude that the photoelectron lifetime is limited by free exciton formation such that $\tau_{e-A^0} \approx \tau_{fX}$. Since the $\mbox{A}^0$X line decays with a similar lifetime, we conclude that the lifetime of excitons bound to acceptors is also limited by free exciton formation, at a rate similar to $1/\tau_{fX}$. Indeed, excitons bound to acceptors have a binding energy of $\sim 2$ meV $< k_BT_e$ with respect to the free exciton state, so that coupling with free excitons should be efficient. 
We expect therefore $\tau_{A^0X}\approx \tau_{e-A^0} \approx \tau_{fX}$ since both populations are coupled with a reservoir of free excitons which we assume have a very short non-radiative lifetime (for example, via formation and escape of exciton-polaritons). 	This value for the exciton formation time ($0.7-0.9$ ns) is consistent with previous measurements on undoped samples \cite{Amo:2006a}. \
 
 On the other hand, at high excitation power, exciton formation is inhibited and the electron lifetime increases with concentration when passing from $\mbox{A}^0$X recombination to b-b recombination. It is seen that the $\mbox{A}^0$X line eventually merges with the  band-to-band recombination emission whose intensity is proportional to $n K_{bb} [n +p_0]$, where the bimolecular coefficient $K_{bb}$ with free holes has been calculated before \cite{dumke1957} and where one assumes that the photohole and photoelectron concentrations are equal because of charge neutrality \cite{cadiz2015c}. In this electron-hole plasma regime, the photoelectron concentration evolves according to $\partial n /\partial t= g - K_0 N_A^0 n  - K_{bb} [n +p_0]n $. This equation  has an analytical solution, given by

\begin{equation} 
n=n_0 \frac{e^{-t/\tau^*} }{1+n_0 K_{bb} \tau^*[1-e^{-t/\tau^*}]}    
\label{eqn}
\end{equation} 
where $1/\tau^* =K_0 N_A^0+K_{bb}p_0$.  The intensity of the luminescence of the two lines is given by 
\begin{equation} 
I(\mbox{e-A}^0)(t)=A K_0 N_A^0 n_0 \frac{e^{-t/\tau^*} }{1+n_0 K_{bb} \tau^*[1-e^{-t/\tau^*}]}      
\label{eqeA}
\end{equation} 

\begin{align}
I(\mbox{b-b})(t)=& AK_{bb} n_0 p_0 \frac{e^{-t/\tau^*} }{1+n_0 K_{bb} \tau^*[1-e^{-t/\tau^*}]}  \nonumber \\ & + A K_{bb} n_0^{2}\frac{e^{-2t/\tau^*} }{\big\{1+n_0K_{bb} \tau^*[1-e^{-t/\tau^*}]\big\}^2} 
\label{eqBB}
\end{align}

While increasing the excitation density, the second term of Eq. (\ref{eqBB}) becomes progressively dominant so that the characteristic decay time of the b-b line should be one half that of the e$-\mbox{A}^0$ one, as indeed observed in our experiments. It is also expected that the departure from an exponential behavior should be stronger for the b-b line than for the $eA_0$ one, which is indeed the case at short delays and high power (see right panel  of Fig.\ref{Fig02}). The bimolecular recombination coefficient $K_{bb}$ at a carrier temperature of $T_e=50$ K is $K_{bb}=4.27 \times 10^{-9}\;\mbox{cm}^{3}/s$ \cite{dumke1957}.  Since $\tau^* \approx 1.6$ ns at the maximum power used, we estimate a characteristic photocarrier concentration of $ \approx  1/(\tau^* K_{bb}) = 1.46 \times 10^{17}\;\mbox{cm}^{-3}$ when exciting at $5$ mW. \

We now focus on the spin dynamics. Figure \ref{Fig03}  shows the transients of the luminescence degree of circular polarization $\mathscr{P}$ for selected excitation powers.  Note that the polarization inmediately after the pulse, of the order of 15$\%$, is smaller than the initial polarization of $25\;\%$ expected for GaAs. This reveals losses of polarization during fast energy relaxation and thermalization of the carriers. The polarization transients are essentially exponential and enable us to determine a well-defined spin relaxation time $T_1$ for both lines, as summarized in the middle panel of Fig.  \ref{Fig04}. $T_1$ is found to decrease significantly with excitation power in the range explored (from $T_1\approx 1$ ns at low power for the $\mbox{A}^0$X line to $T_1 \approx 150$ ps for the b-b line at high densities). Interestingly, $T_1$ is lower for the e-$\mbox{A}^0$ line than for the $\mbox{A}^0$X line at low densities, whereas the same spin relaxation time is found at high power for both lines. We can interpret these findings as spin relaxation induced by scattering between photoelectrons and holes (Bir-Aronov-Pikus mechanism) \cite{bir1975}. The free electron spin relaxation time  is given by \cite{zerrouati1988}

\begin{equation} 
\frac{1}{T_1} = \frac{1}{T_1^0} \left[| \psi(0)|^4 \frac{N_A^-+p}{N_A+p} + \frac{5}{3}(1-\frac{N_A^-+p}{N_A+p} ) \right]
\label{bap}
\end{equation}

\noindent
where $1/T_1^0 = N_A a_B^3 \frac{2 v_e}{\tau_0 v_B} $, $a_B= 1.13 \times 10^{-8}$ m is the exciton Bohr radius, $v_B=\hbar/(\mu_X a_B) = 1.7 \times 10^{5}$ m/s is the exciton Bohr velocity with $\mu_X\approx m_e^*$ the exciton reduced mass and $m_e^*$ the electron's effective mass in the conduction band, $v_e= \sqrt{3k_B T_e/m_e^*}$  is the thermal velocity of electrons, $\tau_0 =\hbar E_B/\Delta^2= 1.1$ ns can be estimated from the exchange splitting  $\Delta \sim 0.05$ meV of the exciton ground state \cite{Fishman:1977a, Ekardt:1979a} and from the exciton binding energy $E_B=4.2$ meV \cite{Nam:1976a}, $p$ represents the density of photoholes, $N_A^-$ the density of ionized acceptors and $|\psi(0)|^2$ is the Sommerfeld factor, estimated to be $|\psi(0)|^2 \approx 2 \pi$ for $T_e=50 $ K \cite{Shinada:1966a,maialle1996}. We estimate $T_1^0 = 3.63$ ns. At low concentrations, $p \ll N_A^- \approx 0.1 N_A$ and therefore $T_1 \approx T_1^0/(39.47 \times 0.1 + 5/3 \times 0.9 ) = 667  $ ps, which agrees remarkably well with the measured spin relaxation time of $690$ ps for the e-$\mbox{A}^0$ line at low power. The spin relaxation time for the $\mbox{A}^0$X is almost twice as long, because in the $\mbox{A}^0$X complex (one electron and two holes) the total electron-hole exchange interaction vanishes since the stable state of $\mbox{A}^0$X involves a hole singlet of the form $\frac{1}{\sqrt{2}}( \Uparrow\Downarrow - \Downarrow\Uparrow ) $. Exchange interaction of the trapped electron with free electrons or free holes may be responsible for the observed relaxation time of $1$ ns.\

At high densities, the measured spin lifetime is the same for both lines, as expected since in this regime excitons are no longer stable. Using the above estimated value for the characteristic photocarrier concentration ($ 1.46 \times 10^{17}\;\mbox{cm}^{-3} \approx N_A ) $ at $5$ mW, we have $ p  \approx   n \gg N_A^- $ and therefore $T_1 \approx T_1^0/(39.47 \times 0.63 + 5/3\times 0.36) = 140 $ ps, which agrees well with the measured value of $T_1 \sim 150$ ps considering that screening has been neglected in Eq.(\ref{bap}). The solid red curve shown in the middle panel of Fig.\ref{Fig04} is the spin lifetime given by Eq. (\ref{bap}) as a function of excitation power, assuming that the photocarrier concentration is linear in power. This equation reproduces well the observed decrease of the spin lifetime for the e-$\mbox{A}^0$ line as the concentration of free carriers increases. Temperature-dependent measurements (see supplementary information) are in agreement with an electron-spin lifetime limited by exchange interaction with free holes.

The results obtained for the present sample are quite different from those obtained for a larger acceptor concentration, in the $ 10^{18}$ cm$^{-3}$ range, which shows little dependence of spin relaxation time and lifetime as a function of excitation power \cite{cadiz2014}, probably due to more efficient screening of the electron-hole interaction.  
Also shown in the bottom panel  of Fig.\ref{Fig04} is the ratio between the spectrally integrated higher energy line (mixture of $\mbox{A}^0$X and b-b) to the e-$\mbox{A}^0$ line inmediately after the pulse. For sufficiently large densities, the photoexcited carrier concentration is higher than the concentration of neutral acceptors, and the ratio becomes greater than $1$.  The dashed line to the left represents the excitation power at which the photogenerated density at $t=0$ equals the theoretical Mott density  in GaAs at $T= 0 $ K ($n_c= 2.8 \times 10^{16}\;\mbox{cm}^{-3}$) \cite{Haug:1984a}, whereas the dashed line to the right correspond to the experimental value of the Mott density in  undoped GaAs ($n_c=1.2 - 1.8 \times 10^{17}\;\mbox{cm}^{-3}$) \cite{Amo:2006a}. Independent investigations have shown that screening of the exciton gas start to occur above $n=1.6 \times 10^{16}$ cm$^{-3}$ \cite{fehrenbach1982}. 
This screening explains the progressive increase of the recombination time and the fact that the two lines are characterized by the same spin relaxation time for an excitation power larger than $50\;\mu$W.

In summary, a systematic study of the carrier and spin recombination dynamics on moderately doped p-type GaAs has been performed at $15$ K. It has been shown that at low densities, electrons can form exciton complexes ($\mbox{A}^0$X) or recombine with holes bound to neutral acceptors. The free electron lifetime is limited by exciton formation and the spin lifetime is larger in exciton complexes than for free electrons since the electron-hole exchange interaction is suppressed when excitons are trapped on neutral acceptors. In this density range, the free electron's spin lifetime is limited by exchange scattering between electrons and trapped holes.
Increasing the carrier concentration causes screening of the electron-hole Coulomb interaction and eventually excitons are no longer stable. As a consequence, the recombination lifetime increases with concentration and the spin lifetime of free electrons decreases by almost one order of magnitude due to accelerated spin relaxation when free photoholes are added to the system. 

\section{Supplementary Material}
See supplementary material for the reflectivity, photoluminescence excitation spectroscopy, and temperature-dependent measurements. 

\section{Aknowledgements}
F.C and P.R thank the grant NEXT No ANR-10-LABX-0037 in the framework of the Programme des Investissements d'Avenir.





\begin{thebibliography}{33}
\bibitem[{not()}]{notetrpl}
\bibinfo{note}{$\tau_{fX}$ does not depend on excitation power since, in the
  regime where no exciton screening takes place, the photohole concentration is
  smaller than $N_A^-$.}
\end{thebibliography}

\end{document}